\PassOptionsToPackage{dvipsnames,svgnames}{xcolor}

\documentclass[a4paper]{jpconf}
\usepackage{iopams}

\usepackage{graphicx}
\usepackage{xcolor}

\usepackage{import}
\usepackage[lofdepth,lotdepth]{subfig} 
\usepackage{bigints}
\usepackage{dsfont}

\newcommand{\al}{\alpha}
\newcommand{\be}{\beta}
\newcommand{\ga}{\gamma}
\newcommand{\Ga}{\Gamma}
\newcommand{\de}{\delta}
\newcommand{\eps}{\epsilon}
\newcommand{\et}{\eta}
\newcommand{\thet}{\theta}
\newcommand{\la}{\lambda}
\newcommand{\ph}{\phi}
\newcommand{\Ome}{\Omega}
\newcommand{\nind}{\noindent}
\newcommand{\mcl}[1]{\mathcal{#1}}
\newcommand{\mfk}[1]{\mathfrak{#1}}
\newcommand{\msf}[1]{\mathsf{#1}}
\newcommand{\idM}{\mathds{1}}
\newcommand{\<}{\langle}
\renewcommand{\>}{\rangle}
\newcommand{\nf}{$\mathcal{N}=4$ SYM}
\newcommand{\dt}[1]{\dot{#1}}
\newcommand{\tl}[1]{\tilde{#1}}
\newcommand{\del}{\partial} 	
\newcommand{\diff}{\mathop{}\!\mathrm{d}}	
\newcommand{\RP}[1]{\mathbb{RP}^{#1}}

\begin{document}

\begin{flushright}
{\small LMU-ASC 66/16}\\
{\small QMUL-PH-16-23}
\end{flushright}

\noindent\textsc{to appear in the Journal of Physics: Conference Series}

\vspace{-1cm}

\title{Tree-level scattering amplitudes from the amplituhedron}

\author{Livia Ferro,$^1$ Tomasz \L ukowski,$^2$ Andrea Orta$^{1,}$\footnote[99]{Speaker at the \emph{7th Young Researcher Meeting}, 24th--26th October 2016, Torino, Italy.} and Matteo Parisi$^3$}

\address{
$^1$ Arnold Sommerfeld Center for Theoretical Physics, Ludwig-Maximilians-Universit\"at, \\
$\;\;$ Theresienstra\ss e 37, 80333 M\"unchen, Germany \\

$^2$ Mathematical Institute, University of Oxford, Andrew Wiles Building, \\
$\;\;$ Woodstock Road, Oxford, OX2 6GG, United Kingdom \\

$^3$ Center for Research in String Theory, School of Physics and Astronomy, \\
$\;\;$ Queen Mary University of London, Mile End Road, London E1 4NS, United Kingdom
}

\ead{andrea.orta@lmu.de}

\begin{abstract} 
A central problem in quantum field theory is the computation of scattering amplitudes. However, traditional methods are impractical to calculate high order pheno\-meno\-logically relevant observables. Building on a few decades of astonishing progress in developing non-standard computational techniques, it has been recently conjectured that amplitudes in planar $\mcl{N}=4$ super Yang-Mills are given by the volume of the (dual) amplituhedron. After providing an introduction to the subject at tree-level, we discuss a special class of differential equations obeyed by the corresponding volume forms. In particular, we show how they fix completely the amplituhedron volume for next-to-maximally helicity violating scattering amplitudes. 
\end{abstract}


\section{Introduction}\label{introduction}

Scattering amplitudes are among the most fundamental quantities in quantum field theory. The current precision frontier has risen and demands highly non trivial theoretical input. Already several decades ago it became clear that amplitudes are generally simpler than one would imagine looking at their traditional expressions in terms of Feynman diagrams. One of the earliest examples of this phenomenon is the celebrated Parke-Taylor formula for the tree-level colour-ordered scattering amplitude of $n$ gluons in a maximally-helicity-violating (MHV) configuration: parametrizing their null momenta via spinor-helicity variables $\la^\al,\tl\la^{\dt\al}$, one finds \cite{Parke:1986gb}
\begin{equation}
A_n^{\text{MHV}}(1^+,\dots,i^-,\dots,j^-,\dots, n^+) = \frac{\<ij\>^4}{\<12\>\<23\>\cdots\<n1\>} \de^4(p) \, ,
\end{equation}
where $\<ab\> \equiv \eps_{\al\be} \la_a^\al \la_b^\be$. Much progress has been achieved in the understanding of this simplicity, especially in the context of $\mcl N=4$ super Yang-Mills (SYM), a four-dimensional superconformal theory which can be thought of as a (maximally) supersymmetric generalization of QCD: indeed, tree-level gluon amplitudes in the two theories coincide and it has been conjectured that \nf{} always provides a specific part of the QCD result, called maximally transcendental. 

The spectrum of \nf{} comprises sixteen massless states organized in a CPT-self-dual supermultiplet: two gluons, eight gluinos and six scalars, all living in the adjoint representation of the colour group $\text{SU}(N)$. In the following we will assume to have performed a colour decomposition, so that we can focus on the kinematics alone. Supplementing spinor-helicity variables with auxiliary Grassmann-odd variables $\et^{\msf A}$ allows us to collect all states in a superfield
\begin{equation}
\Phi = G^+ + \et^{\msf A} \Psi_{\msf A} + \frac12 \et^{\msf A}\et^{\msf B} S_{\msf A \msf B} + \frac1{3!} \et^{\msf A}\et^{\msf B}\et^{\msf C} \eps_{\msf A \msf B \msf C \msf D}\, \bar\Psi^{\msf D} + \frac1{4!} \et^{\msf A}\et^{\msf B}\et^{\msf C}\et^{\msf D} \eps_{\msf A \msf B \msf C \msf D}\, G^- \,.
\end{equation}
Then the amplitude for the scattering of arbitrary states can be extracted from the super\-amplitude $\mcl A$ computed from superfields, accounting for all possible helicity configurations and admitting a decomposition in helicity sectors:
\begin{equation} \label{factorMHV}
\mcl A_n(\{\la_i,\tl\la_i,\et_i\}) = \mcl A_{n,\text{tree}}^{\text{MHV}}\; \mcl P_n(\{\la_i,\tl\la_i,\et_i\}) \,,\quad \mcl A_{n,\text{tree}}^{\text{MHV}} = \frac{\de^4(p)\de^8(q)}{\< 12\>\< 23\>\cdots\< n 1\>}  \, ,
\end{equation}
where $p^{\al\dt\al} = \sum_i \la_i^\al \tl\la_i^{\dt\al} \;,\; q^{\al \msf A} = \sum_i \la_i^\al \et_i^{\msf A}$ are the conserved total momentum and supermomentum and 
\begin{equation}
\mcl P_n = \mcl P_n^{\text{MHV}} + \mcl P_n^{\text{NMHV}} + \mcl P_n^{\text{N$^2$MHV}} + \dots + \mcl P_n^{\text{N$^{n-4}$MHV}} \, .
\end{equation}
Each (next-to-)$^k$MHV contribution $\mcl P_n^{\text{N$^k$MHV}} \equiv \mcl P_{n,k}$ is a monomial in the $\et^{\msf A}$ of order $\mcl O(\et^{4k})$. 

It turns out that spinor-helicity variables are yet not the best ones to make all the remarkable features of \nf{} manifest. This goal is achieved by \emph{twistor variables}: momentum supertwistors $\mcl Z^{\mcl A} = (\la^\al, \tl\mu^{\dt\al},\chi^{\msf A})$ \cite{Hodges:2009hk} -- where $\tl\mu,\chi$ are related to $\tl\la, \et$ respectively -- are particularly suited to expose a hidden, \emph{dual} superconformal symmetry of the model in the planar limit, \textit{i.e.} when the number of colours $N$ has been taken to be large. The interplay of the latter with the ordinary one gives rise to the Yangian $Y\big(\mfk{psu}(2,2|4)\big)$, an infinite-dimensional symmetry leaving tree-level scattering amplitudes invariant \cite{Drummond:2009fd}. Tree-level scattering amplitudes admit a formulation as integrals over Grassmannian manifolds involving twistor variables \cite{ArkaniHamed:2009dn, Mason:2009qx}. For instance, in terms of momentum supertwistors,
\begin{equation} \label{Grassint}
\mcl P_{n,k} = \frac{1}{\text{Vol}(\text{GL}(k))}\bigintssss_\ga  \frac{\diff^{k\times n}\,c_{\al i}}{(1\,2\, \dots\, k)(2\,3\, \dots \,k+1)\dots(n\,1\, \dots\, k-1)} \prod_{\al=1}^k \de^{4|4}\left(\sum_{i=1}^n c_{\al i}\mcl{Z}_i\right) \,.
\end{equation}
Here the integration variables are the entries of a matrix $C$ parametrizing the points in $G(k,n)$ -- the set of $k$-planes through the origin in $n$ dimensions -- and every $(i\,i+1\,\dots\,i+k-1)$ factor represents a \emph{consecutive}, \emph{ordered} $k\times k$ minor of $C$ involving columns $i,i+1,\dots,i+k-1 \; (\text{mod } n)$. Choosing an appropriate integration contour, encircling a subset of the singularities, any $\mcl P_{n,k}^{\text{tree}}$ can be obtained as the corresponding sum of residues. For additional details, we refer the interested reader to the review \cite{Elvang:2015rqa}. 

The connection to Grassmannians has inspired new geometric and even combinatoric methods for studying amplitudes: on the one hand, Hodges observed in \cite{Hodges:2009hk} that next-to-MHV (NMHV) amplitudes are volumes of polytopes in momentum twistor space; later on, the authors of \cite{ArkaniHamed:2012nw} also showed that the residues of $\mcl P_{n,k}$ are in one-to-one correspondence with on-shell diagrams, objects appearing in the so-called \emph{positroid stratification} of the \emph{positive Grassmannian} $G_+(k,n)$: this space is just the restriction of $G(k,n)$ to the matrices whose $k\times k$ ordered minors are all positive. Both ideas combined led to the amplituhedron proposal \cite{ArkaniHamed:2013jha}, which aims at providing a fully geometric picture of the physics of scattering, at least within planar \nf.

In this paper we will give a short introduction to the tree-level amplituhedron, \textit{i.e.} the object computing tree-level scattering amplitudes, focusing on how they are derived from its geometry. The symmetries of the problem will be discussed, with an emphasis on a set of PDEs called \emph{Capelli differential equations}, which will play a special role. Finally we will present a master formula capturing all NMHV scattering amplitudes as integrals over a \emph{dual} Grassmannian and provide a few examples. We will conclude with some outlook.


\section{The tree-level amplituhedron}\label{amplituhedron}

To define the amplituhedron we need to introduce a bosonized version of the external data encoded in momentum supertwistors $\mcl Z_i^{\mcl A}$. 
We define new variables $Z_i^A$, whose components include the regular momentum twistor variables ${z_i^a\equiv(\la_i^\al,\tl\mu_i^{\dt\al})}$, supplemented by a bosonized version of the fermionic components $\chi_i^{\msf A}$: 
\begin{equation}\label{bosonizedZ}
Z_i^A = \left ( 
\begin{tabular}{c}
  $z_i^a$ \\ $ \phi_1^{\mathsf A}\;\chi_{i\mathsf A}$ \\ \vdots\\ $ \phi_k^{\mathsf A}\;\chi_{i\mathsf A}$
  \end{tabular} 
  \right ) \,,\qquad 
  \begin{aligned}
  i &= 1,\dots, n \\ A &= 1,\dots, m+k  \\ a,\mathsf A &= 1, \ldots, m
  \end{aligned} \;.
\end{equation}
Here the $\ph_{\al}^{\msf A}$ are auxiliary Grassmann-odd parameters and the $Z_i^A$ will be called \emph{bosonized momentum twistors}. Moreover, $n$ is the number of scattering particles, $k$ the next-to-MHV degree and $m$ an even number which takes the value four in the physical case.

Now, let us demand that the external data be \emph{positive}, \textit{i.e.} $Z \in M_+(m+k,n)$, where $Z$ is the matrix whose columns are the individual $Z_i$ and $M_+(m+k,n)$ is the set of $(m+k) \times n$ positive real matrices, \textit{i.e.} matrices whose ordered maximal minors are positive:
\begin{equation}\label{positivity}
\eps_{A_1 \dots A_{m+k}}\, Z_{i_1}^{A_1} \cdots Z_{i_{m+k}}^{A_{m+k}} \equiv \< Z_{i_1}\dots Z_{i_{m+k}} \> > 0 \, ,\quad \text{with}\quad 1\leq i_1 < \dots < i_{m+k} \leq n \, .
\end{equation}
The tree amplituhedron is now the space \cite{ArkaniHamed:2013jha} 
\begin{equation}\label{defamplituhedron}
\mfk{A}^{\text{tree}}_{n,k;m}[Z] := \bigg\{ Y \in G(k,m+k) \;\;:\;\; Y_\al^A=\sum_i c_{\al i} Z^A_i  \quad,\quad C = (c_{\al i} )\in G_+(k,n) \bigg\} \,,
\end{equation}
namely a subspace of the Grassmannian $G(k,m+k)$ determined by positive linear combinations of positive external data. The $k$ vectors $Y_\al$ are non-physical and will be eventually eliminated. 

One can canonically define a $(k\cdot m)$-dimensional differential form $\mathbf\Ome_{n,k}^{(m)}$ on $\mfk{A}^{\text{tree}}_{n,k;m}$, demanding that it has logarithmic singularities on all boundaries of the space: this object is called \emph{volume form}. Indeed, top-dimensional forms on any Grassmannian space must take the form \vspace{-0.25cm}
\begin{equation} \label{Omeganotilde}
\mathbf\Ome_{n,k}^{(m)}(Y,Z)=\prod_{\al=1}^k \< Y_1\cdots Y_k \,d^m Y_{\al} \> \; \Ome_{n,k}^{(m)}(Y,Z) \, .
\end{equation}\vspace{-0.25cm}

\nind Here the \emph{volume function} $\Ome_{n,k}^{(m)}$ simply depends on the auxiliary $k$-plane $Y$ and the external data: it is conjectured to compute the volume of yet another space, dual to $\mfk A_{n,k;m}^{\text{tree}}$. There exist several methods to calculate the volume form. The standard one \cite{ArkaniHamed:2013jha} requires us to triangulate the amplituhedron, \textit{i.e.} to determine a set $\mcl T=\{\Ga_a \}$ of $(k \cdot m)$-dimensional cells of $G_+(k,n)$ whose images in the amplituhedron do not overlap and cover it completely.  One then simply adds up the volume forms of all cells, which in turn admit a very simple expression in terms of local coordinates. The downside of this procedure is that $\mcl T$ is not known in general, as the geometry of the (physical) amplituhedron for $k>1$  is not well understood. In \cite{ArkaniHamed:2014dca} another method was suggested, based on demanding regularity of the form everywhere outside of $\mfk A_{n,k;m}^{\text{tree}}$. Lastly, what we will be using is the following integral representation of the volume function \cite{Ferro:2015grk}: 
\begin{equation} \label{omegaintegral}
\Ome_{n,k}^{(m)}(Y,Z) = \bigintssss_\ga \frac{\diff^{k \times n}\,c_{\al i}}{(1\,2\,\dots\,k) (2\,3\,\dots\,k+1)\cdots(n\,1\,\dots\, k-1)} \prod_{\al=1}^k \de^{m+k}(Y_\al - \sum_i c_{\al i}  Z_i) \, .
\end{equation}
The integral is taken over a suitable contour, in full analogy with the Grassmannian integral \eqref{Grassint}. Each residue corresponds to the volume function on a cell of the tree amplituhedron and, in order to get the proper expression for $\Ome_{n,k}^{(m)}$, we need to take an appropriate sum of them.

The tree-level ``amplitude'' for generic $m$ is then calculated by integrating the canonical form in the following way:
\begin{equation}\label{amplitoampli}
\mcl P_{n,k;m}^{\text{tree}}(\mcl Z)= \int \diff^{m\times k}\,\phi \;\; \Ome_{n,k}^{(m)}(Y^*,Z) \, ,
\end{equation}
where we used a $\de$-function to localize  $\Ome_{n,k}^{(m)}(Y,Z)$ on the reference point $Y^* =$\scalebox{1}{$ \bigg( 0_{m\times k} \; \bigg| \; \idM_{k}\bigg)^{\!T}$}.

Integral \eqref{omegaintegral} will be the starting point for our later derivation. In particular, we begin by considering its properties and symmetries in the case of generic $m$, $n$ and $k$. This allows us to write down a set of differential equations satisfied by the volume function \cite{Ferro:2015grk}, which enforce constraints on $\Omega^{(m)}_{n,k}$ and can be fully solved for $k=1$. In all cases the solution admits an integral representation over the Grassmannian $G(k,m+k)$. We term this space the \emph{dual Grassmannian} and stress that it does not depend on the value of $n$, in contrast to \eqref{omegaintegral}. 


\section{Capelli differential equations and NMHV volume forms}

The tree-level Grassmannian integrals \eqref{Grassint} defined in momentum twistor space possess a lot of interesting properties: in particular, they are Yangian invariant. As was shown in \cite{Drummond:2010uq,Korchemsky:2010ut} this symmetry uniquely fixes their form, up to the contour of integration. We presently lack a realization of the Yangian in the context of the amplituhedron, it is however an interesting question to ask whether it is also possible to determine the volume form directly from symmetries. The answer we provide is positive, at least in the NMHV case. For N$^k$MHV amplitudes with $k\geq2$, however, known symmetries of the amplituhedron are not sufficient to completely fix the expression for the volume. 

After introducing collective variables $W_a^A$ -- equal to $Y_a^A$ for $a=1,\dots,k$ and to $Z_{a-k}^A$ for $a=k+1,\dots,k+n$ -- we can show that the integral \eqref{omegaintegral} enjoys the following properties:
\begin{itemize}
\item $\text{GL}(m+k)$ right covariance: 
\begin{equation}\label{invarianceglobal}  
\Ome^{(m)}_{n,k}( Y \cdot g,  Z \cdot g)= \frac{1}{(\det g)^{k}} \,\Ome^{(m)}_{n,k}(Y,Z) \, ,
\end{equation}
for $g \in \text{GL}(m+k)$, where by the right multiplication we mean $( W \cdot g)^{A}_a=\sum_{B}W^B_a g^{\, A}_{B}$.

\item Scaling, \textit{i.e.} $\text{GL}(k)_+\otimes \text{GL}(1)_+ \otimes\dots \otimes \text{GL}(1)_+$ left covariance:
\begin{equation}\label{scalingglobal}
\Ome^{(m)}_{n,k}(h\cdot Y ,\la\cdot Z) = \frac{1}{(\det h)^{m+k}}\, \Ome^{(m)}_{n,k}(Y,Z) \, ,
\end{equation}
for $h \in \text{GL}(k)_+$ and $\lambda=(\lambda_1,\dots ,\lambda_n)\in \text{GL}(1)_+ \otimes\cdots \otimes \text{GL}(1)_+$, where all transformations belong to the identity component of linear groups, $\text{GL}(l)_+=\{ \ell\in \text{GL}(l):\det \ell>0\}$. 
 
\item \emph{Capelli differential equations} on the Grassmannian $G(m+k,k+n)$: for every $(k+1)\times (k+1)$ minor of the matrix composed of derivatives $\frac{\del}{\del W^A_a}$, one can check that
\begin{equation}\label{Capelli}
\det\mathop{\left(\frac{\partial}{\partial W^{A_\nu}_{a_\mu}}\right)_{1\leq \nu \leq k+1}}_{\hspace{1.7cm}1\leq \mu\leq k+1} \Omega_{n,k}^{(m)}(Y,Z) =0\ \, ,\quad
\begin{aligned}
&1\leq A_1 \leq \dots \leq A_{k+1}\leq m+k \\ &1\leq a_1 \leq \dots \leq a_{k+1} \leq k+n
\end{aligned} \; .
\end{equation}
\end{itemize}

\nind Interestingly, the Capelli equations \eqref{Capelli} together with the invariance and scaling properties \eqref{invarianceglobal}, \eqref{scalingglobal} were already studied intensively by the school of Gelfand \cite{MR841131} and also by Aomoto \cite{Aomoto:1975fr, Aomoto:2011hyper}. For the $k=1$ case, relevant for NMHV amplitudes, the solution of the above problem can be inferred from the results presented in \cite{GelfandGraev:1992hyper} and reads
\begin{equation}\label{volume3}
\Ome^{(m)}_{n,1}= \bigintssss_0^{+\infty} \left(\prod_{A=2}^{m+1} \diff s_A\right) \frac{m!}{\left(s\cdot Y\right)^{m+1}} \prod_{i=m+2}^n \thet\left(s\cdot Z_i\right) \,,
\end{equation}
where $\text{GL}(1+m)$ covariance of \eqref{omegaintegral} enabled us to fix $m+1$ variables $\{Z_1,\dots,Z_{1+m}\}=\idM_{1+m}$. Here $s \cdot W_a  := W_a^1 + s_2\,W_a^2  +\ldots+  s_{1+m} W_a^{1+m}$ and the volume function is thus expressed as an integral over a region in the dual Grassmannian $G(1,1+m) = \RP m$.

\nind A few comments are in order. The integrand depends on the number of particles only through the $\theta$-functions, shaping a domain of integration $\mcl D_n^{(m)}$ where no singularities are present. Indeed, 
\begin{equation}\label{nonsingintegrand}
s \cdot Y = s \cdot (c_i Z_i) = c_i \;(s \cdot Z_i) > 0 \, ,
\end{equation}
since $s \cdot Z_i > 0$ and $Y$ is inside the amplituhedron, \textit{i.e.} $c_i>0$. Positivity of the external data implies that  $\mcl D_n^{(m)}$, bounded by the \hbox{$(m-1)$\nobreakdash-dimensional} subspaces $\ell_{Z_i} : s\cdot Z_i = 0$, is convex. Moreover, the choice of frame allowed by the $\text{GL}(1+m)$ covariance implies that eventually we will have to uplift our results to be functions of $(1+m)$\nobreakdash-brackets. Finally, an important feature of our master formula is that it can be evaluated without referring to any triangulation of the integration domain. 


\section{Examples}

We now briefly present a few examples showing that we can reproduce the known results \cite{ArkaniHamed:2010gg}
\begin{align}
\Ome_{n,1}^{(2)} &= \sum_{i=2}^{n-1}\, [1\,i\,i+1] &&\!\!\!,& \![i\,j\,k] &\equiv \mbox{\large $\frac{\< i\,j\,k\>^2}{\< Y \,i\,j\> \< Y \,j\,k\> \< Y \,k\,i\>} $} \label{m2GenVolForm} \, , \\ 
\Ome_{n,1}^{(4)} &= \sum_{i<j}\, [1\,i\,i+1\,j\,j+1] &&\!\!\!,& \![i\,j\,k\,l\,m] &\equiv \mbox{\large $\frac{\< i\,j\,k\,l\,m\>^4}{\< Y \,i\,j\,k\,l\> \< Y \,j\,k\,l\,m\> \< Y \,k\,l\,m\,i\> \< Y \,l\,m\,i\,j\>\< Y \,m\,i\,j\,k\>} $} \label{m4GenVolForm} \,. 
\end{align}
Although \eqref{m2GenVolForm} and \eqref{m4GenVolForm} were obtained triangulating the amplituhedron, we stress that our formula does not rely on this fact at all.
Scattering amplitudes in planar $\mathcal{N}=4$ SYM correspond to $m=4$, nevertheless we find it advantageous to study formula \eqref{volume3} first in the two-dimensional toy model with $m=2$.  

\begin{figure}[h]
\centering
\subfloat[][3-pt]{\label{m2-3pt}
\def\svgwidth{3.25cm}
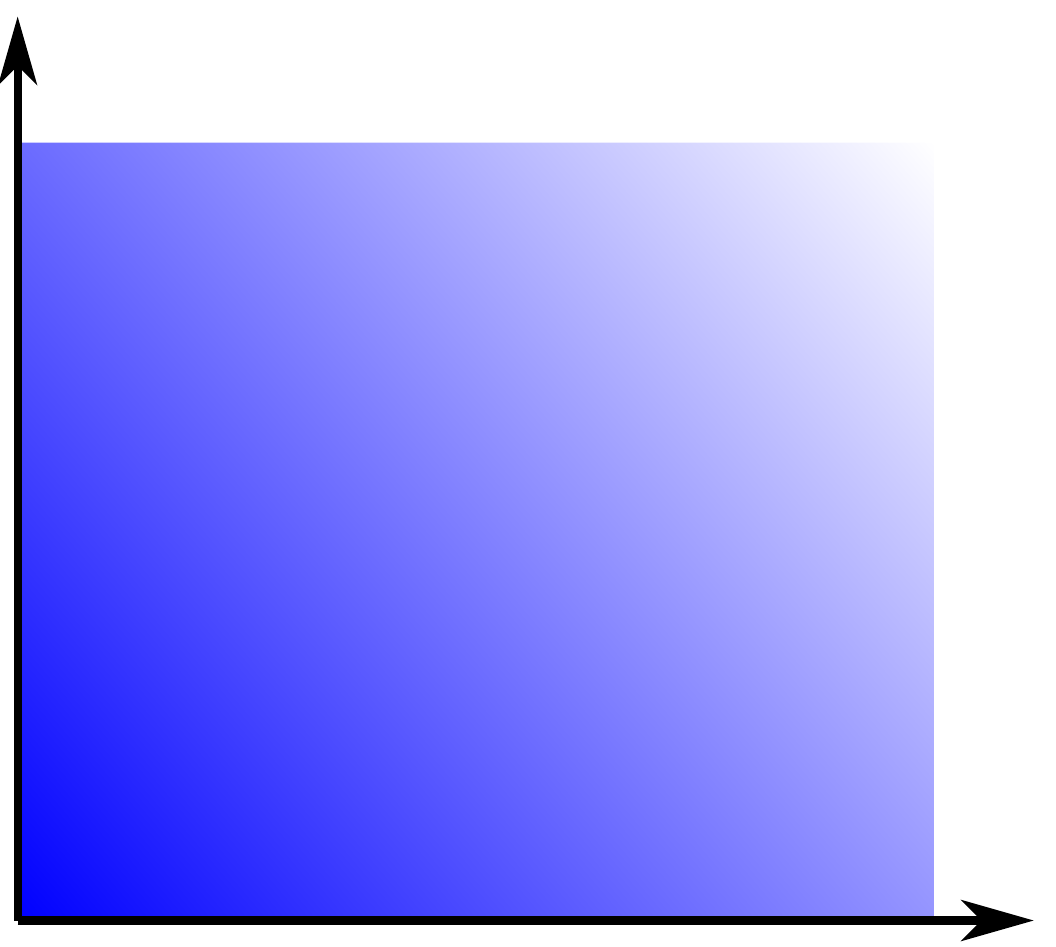} \quad\quad\quad
\subfloat[][4-pt]{\label{m2-4pt}
\def\svgwidth{3.25cm}
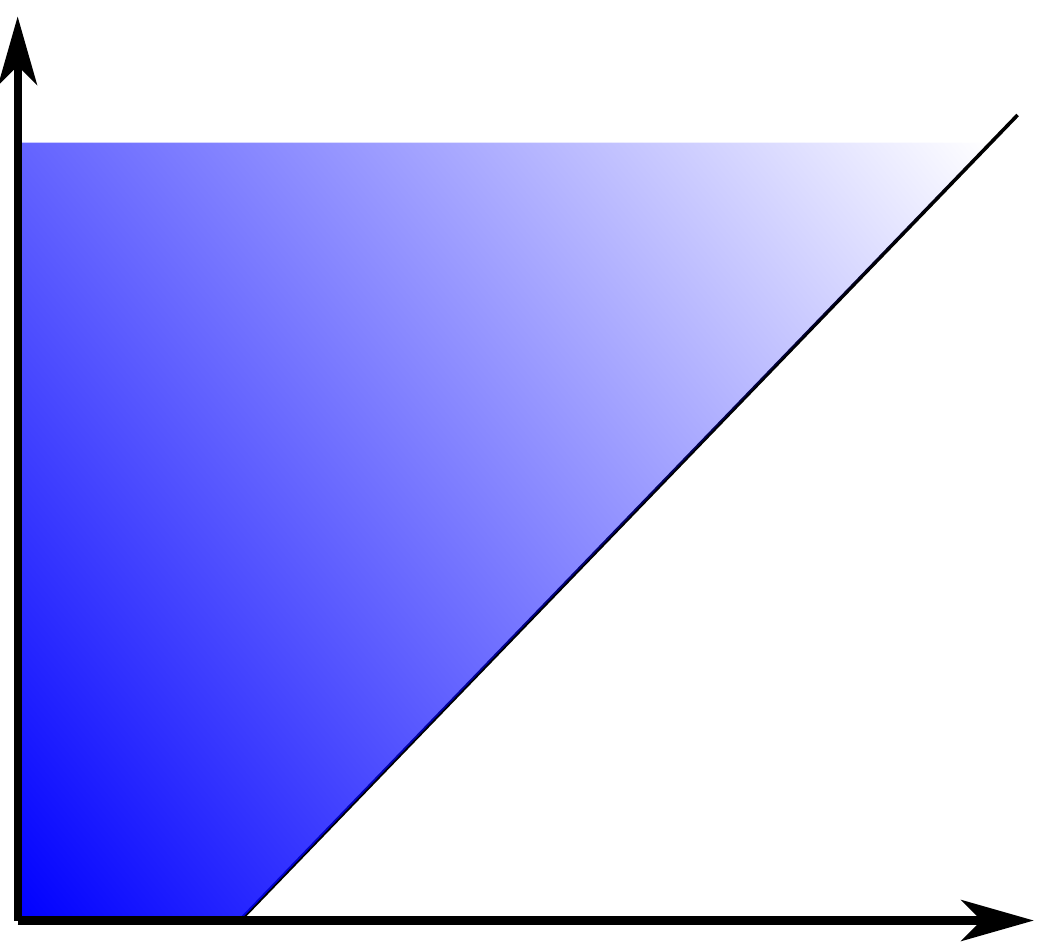} \quad\quad\quad
\subfloat[][$n$-pt]{\label{m2-npt}
\def\svgwidth{3.25cm}
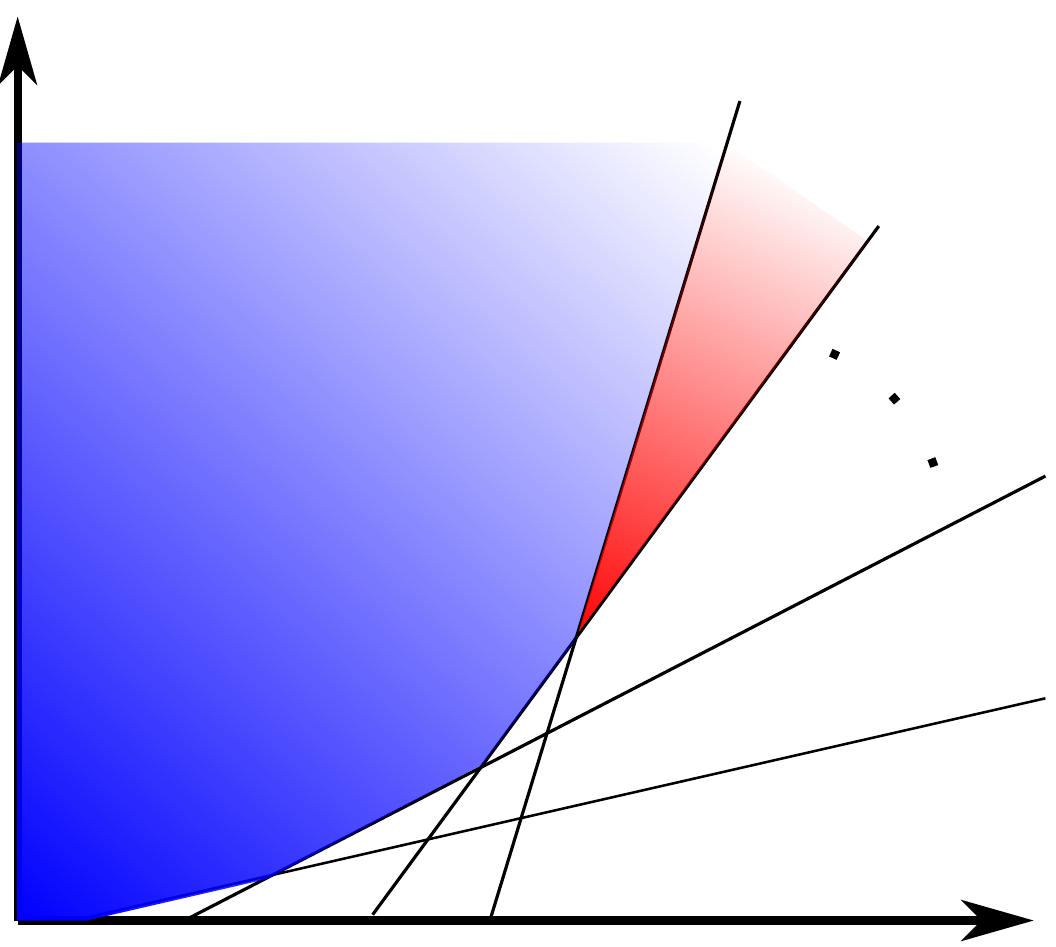}
\caption{The $m=2$ toy model results in a nutshell.}\label{m2-figure}
\end{figure}

At three points there are no $\theta$-functions and the volume function is to be evaluated integrating over the full positive quadrant of the $(s_2,s_3)$ plane: we find $\Ome_{3,1}^{(2)} = (Y^1 Y^2 Y^3)^{-1}$, which correctly lifts to the invariant $[1\,2\,3]$. At four points, $\thet(s \cdot Z_4)$ implies that $\mcl D_4^{(2)}$ is the restriction of $\mcl D_3^{(2)}$ to the region above the line $\ell_{Z_4}$. In general, moving from $n-1$ to $n$ points means one $\thet$-function constraint more, which in turn removes a wedge $\mcl W$ from the integration domain $\mcl D_{n-1}^{(2)}$ (see figure \ref{m2-figure}). Since
\begin{equation}
\int_{\mcl W} \frac{\diff s_2 \diff s_3}{(s \cdot Y)^3} = - [1\,n-1\,n] \, ,
\end{equation} 
the general formula \eqref{m2GenVolForm} holds.

\begin{figure}[]
\centering
\def\svgwidth{0.99\columnwidth}
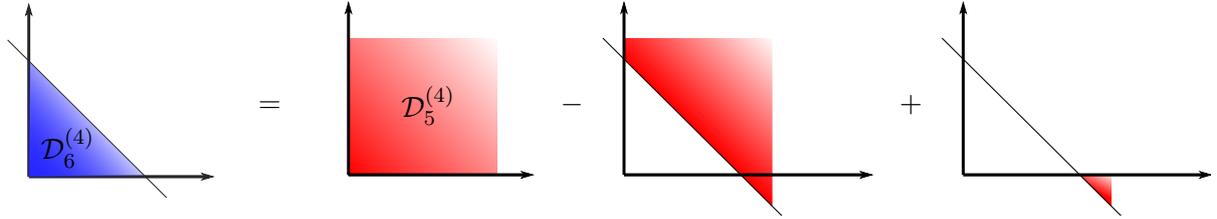
\caption{Domains of integration for six points and $m=4$}\label{m4-6pts}
\end{figure}

\vspace{-0.4cm} 
Moving on to $m=4$, we cannot visualize the four-dimensional integration domains $\mcl D_n^{(4)}$ anymore. The exceptions are the two lowest-$n$ cases, for which it is useful to project everything on the $(s_2,s_4)$ plane. The simplest volume function occurs for five-particle scattering and $\Ome_{5,1}^{(4)} = [1\,2\,3\,4\,5]$, as expected. The six-point calculation is subtler, due to the presence of one $\thet$-function, but still yields the correct result
\begin{equation}\label{m4_6pt}
\Ome_{6,1}^{(4)} = [1\,2\,3\,4\,5] + [1\,2\,3\,5\,6] + [1\,3\,4\,5\,6] \, .
\end{equation}
Remarkably, a convenient integral representation exists for every invariant appearing in equation \eqref{m4_6pt}, allowing for a graphical representation of it, shown in figure \ref{m4-6pts}. Many more details on these derivations can be found in \cite{Ferro:2015grk}.
%


\section{Discussion and outlook}

In this paper we gave a brief introduction to the amplituhedron idea, a recently proposed approach to the calculation of planar scattering amplitudes of \nf. We studied the symmetries of the problem at tree-level and explained how they lead to the novel formula \eqref{volume3}, relevant for NMHV amplitudes. These can now be computed without any reference to triangulations of the amplituhedron. Rather, they admit an integral representation over a dual Grassmannian space $G(1,1+m)$. This suggests a natural generalization to higher-$k$ amplitudes, leading to a framework where the volume functions are integrals over the dual Grassmannian $G(k,m+k)$. For $k>1$, though, symmetry constraints are not enough to fix the final formula completely. Further studies are needed and it would be particularly interesting to understand whether Yangian symmetry could be directly realized in the bosonized momentum twistor space. 


\section*{Acknowledgements}
A.O. is grateful to the organizers of the \textit{7th Young Researcher Meeting} for giving him the opportunity to present and publish his work. L.F. is supported by the Elitenetwork of Bavaria. T.L. is supported by ERC STG grant 306260. The work of M.P. is funded by QMUL Principal's Studentship. 

\section*{References}
\bibliographystyle{iopart-num}
\bibliography{YRM_bibliography.bib}

\end{document}